\documentclass[]{spie}  

\usepackage{amsmath,amsfonts,amssymb}
\usepackage{graphicx}
\usepackage[colorlinks=true, allcolors=blue]{hyperref}
\usepackage{txfonts}
\usepackage{textcomp}
\usepackage{multirow}
\usepackage[table,xcdraw]{xcolor}
\usepackage{lscape}

\title{Polarimetric modeling and assessment of science cases for Giant Magellan Telescope-Polarimeter (GMT-Pol)}

\author[a]{Ramya M Anche}
\author[a]{Grant Williams}
\author[c]{Hill Tailor}
\author[b]{Chris Packham}
\author[a,c]{Daewook Kim}
\author[a,c]{Jaren N Ashcraft}
\author[a]{Ewan S. Douglas}
\author[*]{GMT-Pol team}
\affil[a]{Steward Observatory, University of Arizona, 933N Cherry Avenue, Tucson, Arizona, 85721, USA}
\affil[b]{University of Texas, San Antonio, 1 UTSA Circle, San Antonio, TX 78249}
\affil[c]{James C. Wyant College of Optical Sciences, University of Arizona, 933N Cherry Avenue, Tucson, Arizona, 85721, USA}

\authorinfo{Further author information: (Send correspondence to R.M.A.)\\R.M.A.: E-mail: ramyaanche@arizona.edu}

\begin{document} 
\maketitle

\begin{abstract}
Polarization observations through the next-generation large telescopes will be invaluable for exploring the magnetic fields and composition of jets in AGN, multi-messenger transients follow-up, and understanding interstellar dust and magnetic fields. The 25m Giant Magellan Telescope (GMT) is one of the next-generation large telescopes and is expected to have its first light in 2029. The telescope consists of a primary mirror and an adaptive secondary mirror comprising seven circular segments. The telescope supports instruments at both Nasmyth as well as Gregorian focus. However, none of the first or second-generation instruments on GMT has the polarimetric capability. This paper presents a detailed polarimetric modeling of the GMT for both Gregorian and folded ports for astronomical B-K filter bands and a field of view of 5 arc minutes. At 500nm, The instrumental polarization is 0.1\% and 3\% for the Gregorian and folded port, respectively. The linear to circular crosstalk is 0.1\% and 30\% for the Gregorian and folded ports, respectively. The Gregorian focus gives the GMT a significant competitive advantage over TMT and ELT for sensitive polarimetry, as these telescopes support instruments only on the  Nasmyth platform. We also discuss a list of polarimetric science cases and assess science case requirements vs. the modeling results. Finally, we discuss the possible routes for polarimetry with GMT and show the preliminary optical design of the GMT polarimeter. 
\end{abstract}

\section{Introduction}
\label{sec:intro}  
Astronomical polarimetry has played a significant role in understanding various general astrophysical phenomena, from galactic magnetic fields around interstellar dust to synchrotron radiation from active galactic nuclei \cite{clarke2009stellar}. In the field of exoplanets and circumstellar disks, polarimetry in conjunction with high contrast imaging not only reveals the asymmetrical structure of the disks, scattering by dust grains and refractive indices of aerosol and molecules in the exoplanets' atmosphere, it also improves suppression of the unpolarized star point spread function (PSF). The degree of polarization \cite{leroy200pol}, which is of interest to astronomers, ranges from several tenths of a percent to as low as $10^{-5}$. 
\par As polarimetry is a photon-starved technique, with the next-generation of 30m class telescopes such as the Extremely Large Telescope \cite{cayrel2012elt} Thirty Meter Telescope \cite{nelson2006tmt},  and Giant Magellan telescope \cite{bernstein2014overview}, one can aim to obtain high accurate polarimetric data of the objects fainter than $V=20$. However, one significant challenge for accurate polarimetry is the polarization changes introduced by the telescope optics to the incoming polarization \cite{tinbergen2005astronomical}. The instrumental polarization (polarization introduced to the unpolarized light) and cross-talk (conversion from linear to circular polarization or vice versa) due to the telescope optics of the Cassegrain telescope are found to be on the order of $0.1\%$ and increase to a few percent for Nasmyth telescopes \cite{keller2002instrumentation}. For some of the existing and future polarimetric instruments, these effects are modeled using polarization ray tracing algorithms\cite{chipman1995mechanics} to understand the nature and variation of these effects and design efficient calibration and mitigation strategies \cite{harrington2017polarization,van2020polarimetric,harrington2011deriving,sen1997instrumental,maharana2022walop}. 
\par In the context of the next generation, giant segmented mirror
telescopes (GSMTs), the polarization effects from telescope optics of only the TMT and ELT have been evaluated and presented. The polarization modeling for the Thirty Meter Telescope (with monolith primary) by \textit{Anche et al.} \cite{anche2018analysis,atwood2014polarimetric} showed the variation of instrumental polarization of 4.5-0.6\% and crosstalk of 73-11\% for wavelengths 0.4-2.5 $\mu$m for an on-axis star. Similarly, for ELT with the monolith primary mirror, \textit{de Juan Ovelar et al.} estimated an instrumental polarization and crosstalk of 6\% and 30\%, respectively, for zenith angle of 0\textdegree and wavelength of 0.55 $\mu$m \cite{de2014instrumental}. \textit{Anche et al.} \cite{anche2023polarization} has also presented the polarization aberrations of all three GSMTs in the context of High-contrast imaging of exoplanets. This paper presents the polarization modeling of the Giant Magellan Telescope, a list of polarimetric science cases, and derived technical requirements. We also provide the possible routes for integrating polarimetry (GMT-Pol) with the current and proposed instruments for GMT. 
\par
Section \ref{sec1} briefly describes polarimetric science cases and technical requirements for the GMT-Pol. The polarimetric modeling and its results, Mueller matrix, instrumental polarization, and crosstalk, are presented in  Section \ref{sec2}. The polarization effects of all the GSMTs are compared in Section \ref{sec3}. Section \ref{sec4} presents the different possible routes of polarimetry and preliminary design with the Commissioning Camera (ComCam)\cite{crane2020conceptual}. Finally, the conclusions and future work are in Section \ref{sec5}
\section{GMT polarimetry team and Description of Science cases}
\label{sec1}
GMT Polarimetry modeling team was formed in March 2023, consisting of 25 scientists from all over the world, to prepare a scientific justification and technical requirement white paper for the GMT-Polarimeter (GMT-Pol). The six major science areas focused on in the white paper (\textit{Williams et al. 2023 in preparation}) are:
\begin{itemize}
    \item Transients: Novae, Normal and Superluminous Supernovae, Gamma Ray Bursts, Gravitational Wave sources/Kilonovae, Tidal Disruption Events.
    \item Stars and their environments: Stellar wind accretion in symbiotic stars.
    \item Galaxies: Active Galactic Nuclei, Weak lensing and optical polarization of galaxies, Dust in High Redshift Galaxies, Polarization of Lyman Alpha Nebulae and CGM. 
    \item Solar system objects: Regolithic surfaces, Planetary aerosols, Cosmic dust, Magnetic fields, and Astrobiology
    \item Interstellar and circumstellar media: RAT Alignment in a Nutshell, Magnetic Fields in ISM and CSM Environments, Dust and ISM/CSM Environment Parameters, Line Polarization by Ground-State Alignment
    \item Exoplanets and Circumstellar disks: Exoplanets, Protoplanetary, Transition, and Debris disks. 
\end{itemize}

For many of the science cases in the white paper, the targets of interest are too faint to achieve the necessary signal-to-noise to discriminate between models, even with today's largest telescopes and their existing polarimetric capabilities.  Measuring signals at the 0.1\% level (i.e., \ 1 part in 1000) requires the collection of approximately $1000^2$ or 1,000,000 photons.  When trying to achieve that level of sensitivity with spectropolarimetry, a million photons are needed per spectral resolution element.  The Giant Segmented Mirror Telescopes (GSMTs) will provide the increased light-gathering power that will enable an unprecedented era of discovery with astronomical polarimetry.  An observation that would take 14 hours on an 8.2-m telescope can be done in 1.5 hours on the GMT.  Below we provide just a few examples of science cases that would be advanced in a transformative way with polarimetry on a GSMT.

The physical mechanisms that are believed to power many astrophysical transients are not spherically symmetric.  Some examples include merging neutron stars that produce gravitational waves, tidal disruption events in which a star is torn apart by a supermassive black hole, and the highly collimated gamma-ray burst sources known as hypernovae.  The models for all of these phenomena have a preferred axis and, as such, likely have a non-zero polarization that depends on the level of asymmetry and the viewing angle.  Detecting these sources using different techniques (i.e., gravitational waves, gamma-rays, X-rays, or radio) has become common, but their optical counterparts are usually quite faint.  Many astrophysical transients, such as superluminous supernovae, are intrinsically very bright but are extremely rare and therefore are most likely to be detected at very large distances.  The resulting faint optical transients often get fainter with time and require large telescopes to achieve the necessary signal-to-noise.

Other science cases require observations of very faint targets.  For example, for Ly$\alpha$ nebulae, polarimetry provides a key discriminator between the possible sources of either photoionization throughout the nebulae or scattering of Ly$\alpha$ photons produced at the center of Ly$\alpha$ halo. Currently, polarimetry on 8m-class telescopes is limited only to the brightest Ly$\alpha$ nebulae and even then with significant telescope time (20-40h). GMT will enable a systematic survey of high-z Ly$\alpha$ nebulae and constrain the properties of CGM and cosmic web at $z = 2 - 6$.

For many of the science cases in the white paper, the interstellar polarization (ISP) must be determined prior to understanding the intrinsic polarization of the source.  However, for the study of the interstellar medium and galactic magnetic fields, the ISP is the source. The interstellar medium is extensive but can vary dramatically over small distances.   The true scale of the ISM may be even smaller than current surveys can resolve.  Also, multiple clouds along the line of sight can significantly change the polarization spectrum.  To utilize the powerful tool of spectropolarimetry to probe these structures, we, therefore, need to be able to select our background stars with high-spatial accuracy in 3D or at higher extinction, implying targets that are fainter than what's possible with today's capabilities.

For each science area, the instrument requirements, such as wavelength coverage, field of view, and spectral/spatial resolution, required SNR, in addition to the requirements on acceptable instrumental polarization and crosstalk, are described. A brief list of these science cases and their technical requirements are shown in Table \ref{science-areas}. Comparing different observational requirements shows that most of the polarimetric observations require Linear Stokes measurement ($Q$ and $U$) using a moderate resolution spectropolarimeter in the wavelength range of 0.35-1.0 $\mu$m with the field of view of 1-2 arc minutes with the acceptable level of instrumental polarization of $\sim$ 0.05\% and crosstalk of $\sim$ 0.1\%. In the next section, we describe the polarimetric modeling of the GMT telescope to estimate the instrumental polarization and crosstalk.
\section{Modelling of polarization effects from Giant Magellan Telescope optics}
\label{sec2}
\begin{figure}[!h]
    \centering
    \includegraphics[width=0.95\textwidth]{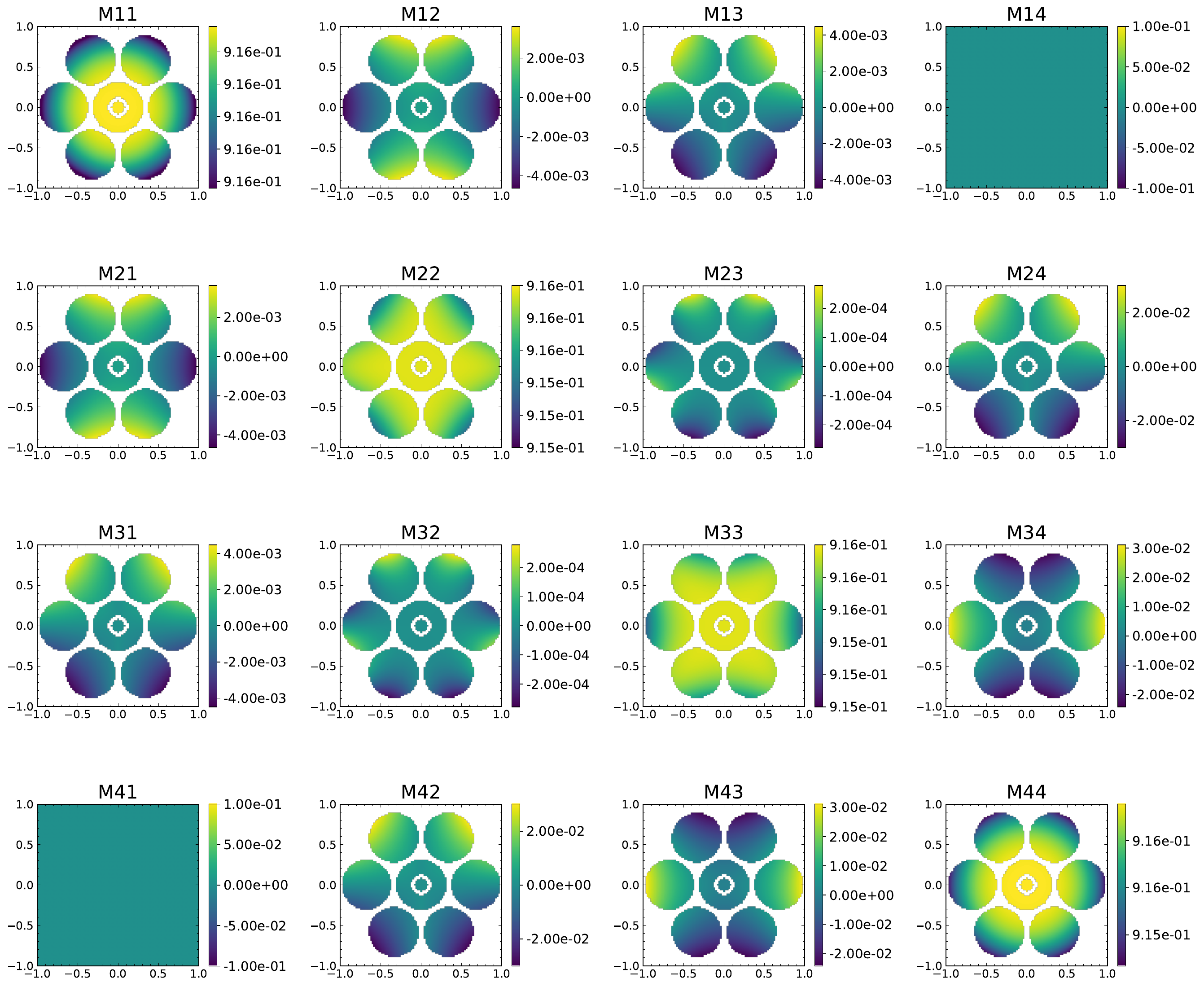}
    \caption{The Mueller matrix of the primary mirror of GMT for $V$ band for the on-axis rays. $M11$, $M22$, $M33$, and $M44$ show the azimuthal symmetry with values increasing from the center to the mirror's edge. The azimuth anti-symmetry is seen in all the elements except the diagonal elements, and $M14$ and $M41$ are found to be zero. For $M21$ and $M31$ contributing to the instrumental polarization, one quadrant of the mirror exhibits positive values, which gets compensated by the other quadrant, which results in negligible values. Similar behavior is also seen for the crosstalk terms ($M43$, $M34$, and $M42$). $x$ and $y$ axis in each panel correspond to normalized pupil coordinates ($px$ and $py$).}  
\label{fig: Mueller-pri}
\end{figure}

The polarization effects from the telescope and instrument optics are estimated using a polarization ray tracing algorithm described in detail in \cite{anche2018analysis,chipman1995mechanics,harrington2011deriving}. Here, we perform the ray tracing of 10000 rays through the telescope mirrors using Zemax Optics studio and estimate the polarization effects and telescope Mueller matrices in Python. We use the optical design of GMT with seven primary mirror segments of size 8.365m, a secondary mirror of 7 segments of size 1.05m, and a Nasmyth mirror of size 0.3m in diameter. The entire list of parameters of the optical design is provided in \cite{anche2023polarization}. GMT will have instrument mounting in both Gregorian and Nasmyth focus locations, for which we estimate the polarization effects using bare Aluminum coating on all three mirrors. Figure \ref{fig: Mueller-pri} shows the Mueller matrix of the primary mirror on the normalized pupil coordinates. $M21$ and $M31$ indicate the instrumental polarization, $M42$, $M43$, and $M34$ indicate the linear to circular crosstalk. The diagonal elements ($M11$,$M22$,$M33$,$M44$) correspond to $I\rightarrow I$, $Q\rightarrow Q$, $U\rightarrow U$, $V\rightarrow V$ respectively. The diagonal elements follow the variation of reflection coefficients $r_p$ and $r_s$ on the mirror surface. The elements corresponding to the instrumental polarization and crosstalk show positive and negative values in consecutive quadrants, which result in negligible values when averaged for all the on-axis rays at the primary focus. A similar Mueller matrix is obtained after the Gregorian secondary of GMT. 
\begin{landscape}
\begin{table}[!h]
\centering
\caption{ A brief list of polarimetric science programs, and observing requirements and acceptable instrumental polarization and crosstalk. The complete description of all the science areas is provided in detail in (\textit{Williams, G et al. 2023 in preparation})}
\label{science-areas}
\footnotesize
\begin{tabular}{llllllllllll}
\hline
{\color[HTML]{000000} Science areas}                                          & {\color[HTML]{333333} \begin{tabular}[c]{@{}l@{}}Wavelength \\ ($\mu$m)\end{tabular}} & {\color[HTML]{000000} FOV} & {\color[HTML]{333333} \begin{tabular}[c]{@{}l@{}}Type of \\ observation\end{tabular}} & {\color[HTML]{000000} \begin{tabular}[c]{@{}l@{}}Linear/\\ Circular\end{tabular}} & {\color[HTML]{000000} \begin{tabular}[c]{@{}l@{}}Target \\ brightness\end{tabular}} & {\color[HTML]{000000} \begin{tabular}[c]{@{}l@{}}Spectral \\ resolution\end{tabular}} & {\color[HTML]{000000} \begin{tabular}[c]{@{}l@{}}Spatial \\ resolution\end{tabular}} & {\color[HTML]{330001} \begin{tabular}[c]{@{}l@{}}Req \\ SNR\end{tabular}} & {\color[HTML]{000000} \begin{tabular}[c]{@{}l@{}}Level of \\ pol per \\ spectral/\\ spatial \\ element\end{tabular}} & {\color[HTML]{000000} \begin{tabular}[c]{@{}l@{}}Acc \\ IP\end{tabular}} & {\color[HTML]{000000} \begin{tabular}[c]{@{}l@{}}Acc \\ CT\end{tabular}} \\ \hline \\
Transients                                                                    & 0.35-0.9                                                                              & 1’                         & \begin{tabular}[c]{@{}l@{}}Moderate resolution \\ Spectro-polarimetry\end{tabular}    & Linear                                                                            & 10-19th                                                                             & 500-1000                                                                              & \begin{tabular}[c]{@{}l@{}}Seeing \\ limited\end{tabular}                            & \begin{tabular}[c]{@{}l@{}}10 at 19th\\ mag\end{tabular}                       & 0.1\%                                                                                                                         & 0.05\%                                                                          & 0.05\% \\
\begin{tabular}[c]{@{}l@{}}Stars and\\ their environments\end{tabular}        & 0.37-0.72                                                                             & 0.5'                       & \begin{tabular}[c]{@{}l@{}}Moderate resolution \\ Spectro-polarimetry\end{tabular}    & Linear                                                                            & -                                                                                   & \textgreater{}10000                                                                   & -                                                                                    & -                                                                              & -                                                                                                                             & -                                                                               & -                                                                                      \\
Galaxies                                                                      & 0.36-0.9                                                                              & \textgreater{}1’           & \begin{tabular}[c]{@{}l@{}}Imaging/ Spectro/\\ Ideally IFU\end{tabular}               & Linear                                                                            & -                                                                                   & 100                                                                                   & 1’’ or better                                                                        & 5$\sigma$                                                                      & 10-20\%                                                                                                                       & \textless{}0.1\%                                                                & -                                                                                      \\
Solar system objects                                                          & 0.4-1                                                                                 & 1-2'                       & Imaging                                                                               & \begin{tabular}[c]{@{}l@{}}Linear+\\    \\ Circular\end{tabular}                  & \begin{tabular}[c]{@{}l@{}}-12 (Moon)\\    \\ -24 (KBO)\end{tabular}                & \textgreater{}500                                                                     & -                                                                                    & -                                                                              &                                                                                                                               & 0.01\%                                                                          & \textless{}1\%                                                                         \\
\begin{tabular}[c]{@{}l@{}}Exoplanets and \\ circumstellar disks\end{tabular} & 0.4-10                                                                                & 1'                         & High-contrast imaging                                                                 & Linear                                                                            & -                                                                                   & -                                                                                     & -                                                                                    & -                                                                              & 10-30\%                                                                                                                       & \textless{}1\%                                                                  & \textless{}1\% \\ \hline                                                                       
\end{tabular}
\end{table}
\end{landscape}
\subsection{Instrumental Polarization (IP) and Crosstalk (CT) at the Gregorian focus}
\begin{figure}[!h]
    \centering
    \includegraphics[width=1\textwidth]{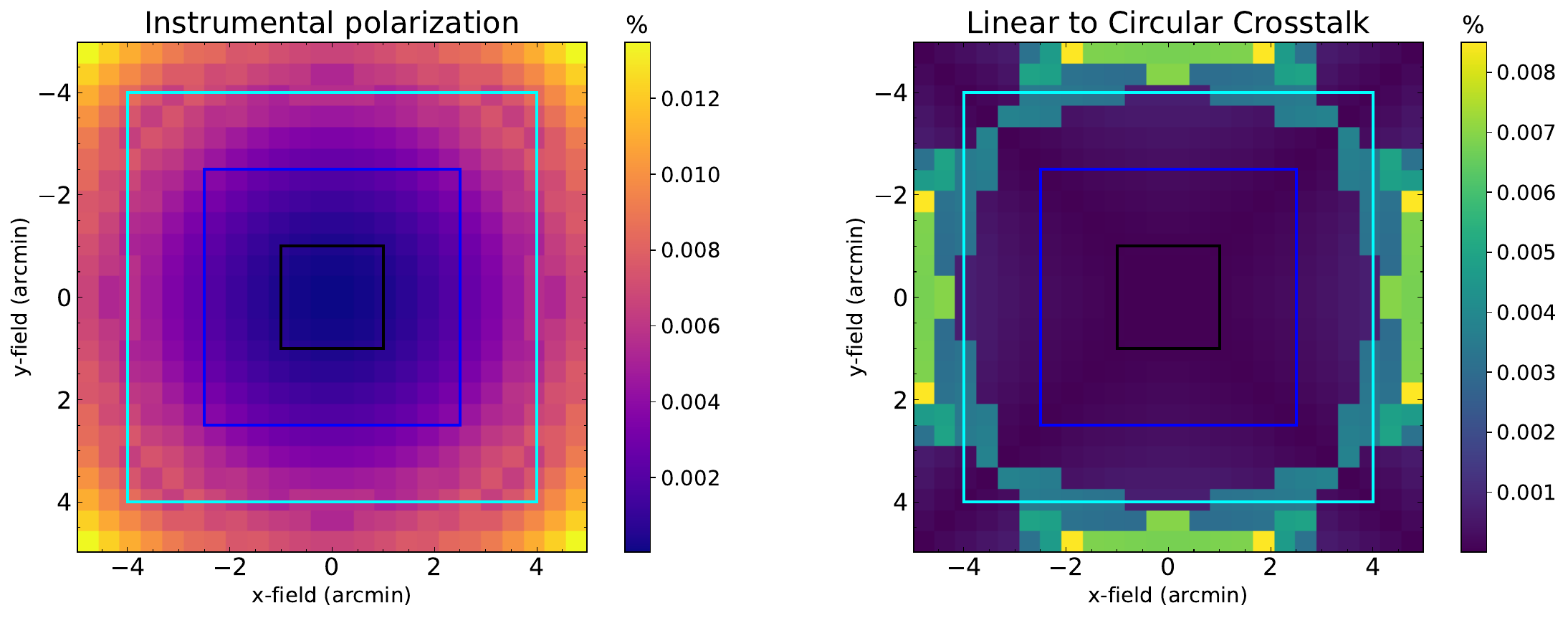}
    \caption{Instrumental polarization and crosstalk estimated at the Gregorian focus of GMT for 5' field of View at 0.54$\mu$m. The three squares correspond to 1', 2.5', and 4' FOV. }
    \label{fig:polgreg}
\end{figure}

\begin{figure}[!h]
    \centering
    \includegraphics[width=1\textwidth]{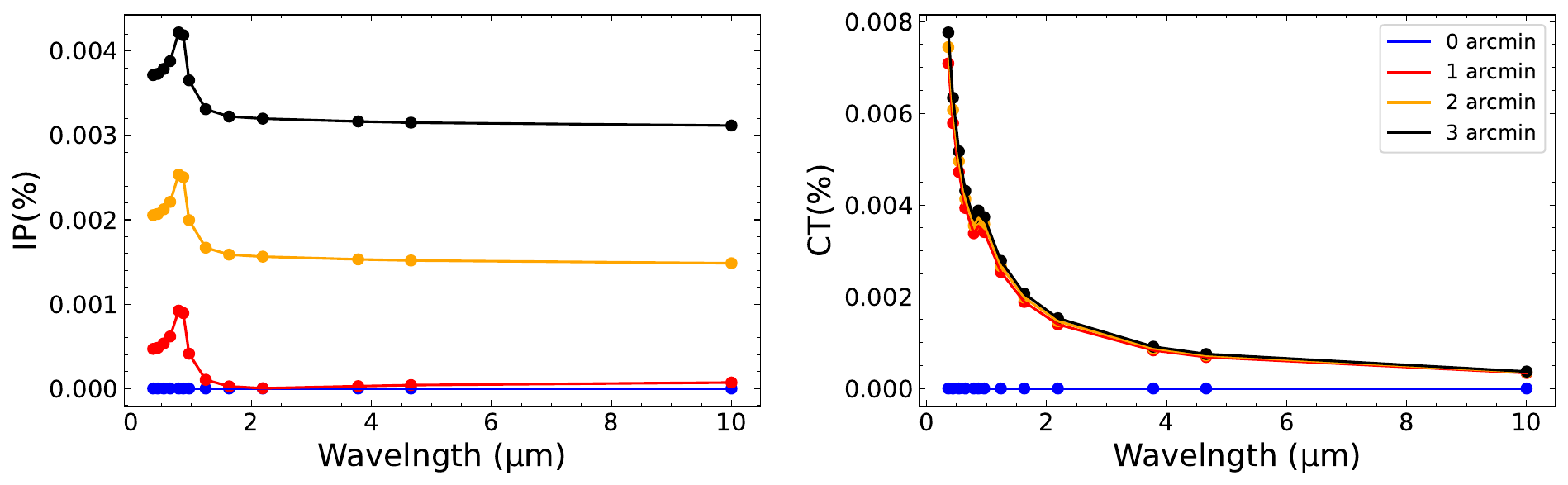}
    \caption{Variation of Instrumental polarization and crosstalk estimated at the Gregorian focus of GMT for wavelengths 0.35$\mu$m to 10$\mu$m.}
    \label{fig:polgregfield}
\end{figure}

Using the polarization ray tracing algorithm, we estimate the Mueller matrices for a field of view of 5' at a wavelength of 0.54$\mu$m. The variation of IP and CT over the 5' FOV, divided into 24 $\times$ 24 arrays of field positions, is shown in Figure \ref{fig:polgreg}. The IP and CT increase with field angle as the asymmetry increases. For FOV $\sim$ 4', the IP and CT values are estimated to be $<$ 0.01\%, which agrees with the requirements specified in the science cases. Figure \ref{fig:polgregfield} shows the variation of IP and CT over the wavelength of 0.35-10$\mu$m for different field angles. The IP values increase in the optical and near IR wavelength region and decrease in the near IR and mid-IR wavelengths, which is directly related to the variation of the refractive index of Aluminum. The CT values decrease with the wavelength and show a small bump near 0.85$\mu$m (at the position where IP shows the maximum value). 
\section{Comparison with other GSMTs at the Nasmyth focus}
\begin{figure}[!h]
\centering
    \includegraphics[width=0.44\textwidth]{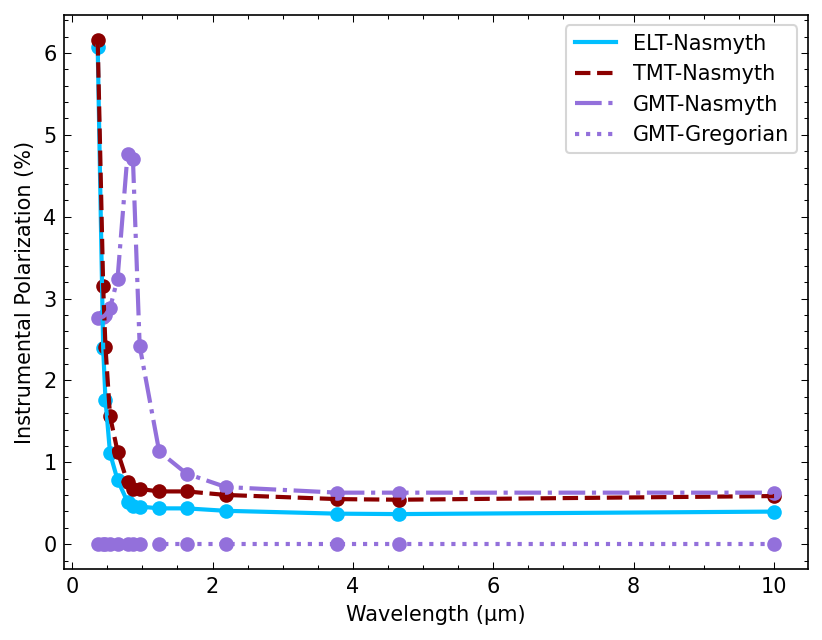}
    \includegraphics[width=0.45\textwidth]{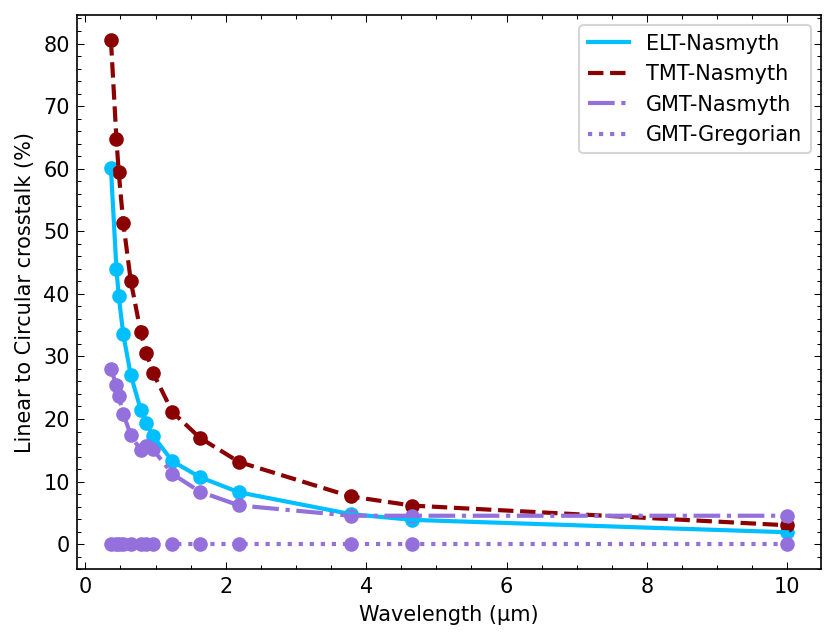}
    \caption{Variation of IP and Crosstalk with wavelength for all the three GSMTs. We use Gemini coating (Ag+$\rm Si_3N_4$) on all the mirrors of ELT and TMT, and bare Aluminum for GMT}
    \label{fig: NasmythfocusGSMTs}
\end{figure}

We estimated the Mueller matrices for all three GSMTs at the Nasmyth focus using polarization ray tracing. Figure \ref{fig: NasmythfocusGSMTs} shows the variation of IP and crosstalk with wavelength. Among the three GSMTs, TMT and ELT have similar IP varying from 6\% to 0.6\% in the wavelength range of 0.4-10 $\mu$m. Although GMT has comparatively less IP than the other two in the optical region, it increases to $\sim$ 5\% in the near-IR region due to the refractive index variation of Aluminum. In the Linear to Circular crosstalk case, TMT shows the highest crosstalk, followed by ELT and GMT. Thus at the Nasmyth port instrument location, GMT will have very similar polarization effects as the other two GSMTs. The advantage of GMT is the availability of Gregorian focus instrument location where the IP and crosstalk are within the requirements for the science cases, as shown in Figure \ref{fig: NasmythfocusGSMTs}. Considering the complexities of polarization effects mitigation and the development of calibration strategies, it is favorable to design a polarimeter for the GMT at the Gregorian focus.  
\label{sec3}
\section{Conceptual design of GMT-Pol}
\label{sec4}
Before the design of GMT-Pol, we intend to investigate the possibility of integrating the polarimetric capability into one of the existing or future instruments for GMT. The first-generation instruments for GMT are:
\begin{itemize}
    \item Giant Magellan Telescope Consortium Large Earth Finder (G-CLEF) \cite{szentgyorgyi2016gmt}- High resolution, high precision radial velocity multi-object spectrograph operating in the range 0.35-0.95 $\mu$m.
    \item Giant Magellan Telescope Multi-object Astronomical and Cosmological Spectrograph (GMACS) - Multi-object medium resolution wide field spectrograph operating in the range of 0.32-1.0 $\mu$m \cite{depoy2014update}
    \item Giant Magellan Telescope Integral-Field Spectrograph (GMTIFS) \cite{sharp2016gmtifs} - A diffraction-limited spectrograph operating in the range 0.9-2.5 $\mu$m
    \item Giant Magellan Telescope Near-IR Spectrograph (GMTNIRS) \cite{jaffe2016gmtnirs} - Single object Echelle spectrograph operating in the 1.1-5.4 $\mu$m. 
    \item Commissioning Camera (ComCam) \cite{crane2020conceptual}  - Alignment and image quality assessment instrument with 6' FOV operating in the range of 0.36-0.95 $\mu$m.
\end{itemize}
\begin{figure}[!h]
    \centering
    \includegraphics[width=1\textwidth]{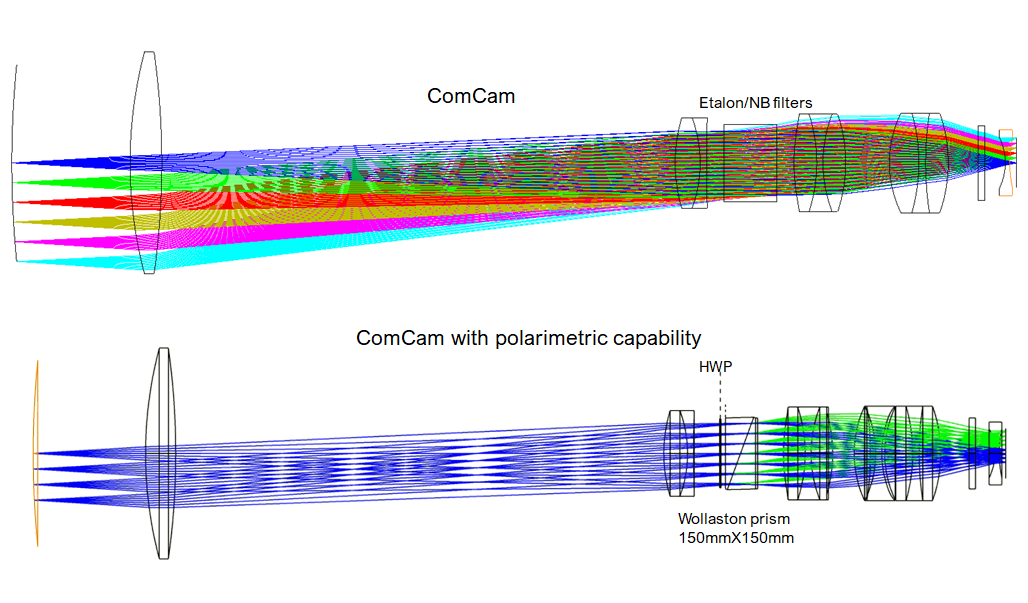}
    \caption{The top panel shows the current ComCam design \cite{crane2020conceptual}, and the bottom panel shows the ComCam with the polarimetric capability. The first two collimator lenses form the pupil at the location of the Etalon/narrow band (NB) filter. To convert it to an imaging polarimeter, a wollaston+HWP can be added at the location of Etalon.}
    \label{fig:comcam-pol}
\end{figure}
Among these, GMTIFS and GMTNIRS operate only in the near/mid-IR wavelength region, and both are fed by the tertiary mirror that introduces significant instrument polarization and crosstalk.  GCLEF, the first instrument on GMT, is a fiber-fed instrument and is already in the stage of development and testing. However, ComCam is currently in the conceptual design phase; the field of view and wavelength range of ComCam overlaps with the technical requirements of the polarimetric science cases. Thus we are currently performing a trade-off study to integrate the polarimeter with the existing ComCam design. The trade-off study aims to analyze the imaging performance after introducing a Wollaston and an HWP in the ComCam optical design without modifying the existing camera and collimator optics. The top panel in Figure \ref{fig:comcam-pol} shows the current ComCam optical design \cite{crane2020conceptual}. The first field lens and the cemented doublet form a nearly collimated beam at the location of the Fabry-Perot Etalon/Narrowband filters. The camera consists of two cemented triplets and a field flattener. The broad-band filter can be inserted in the beam before the field flattener. In this design, we propose to add the Wollaston prism and an HWP at the location of the NB filter in the pseudo-collimated beam and analyze the image quality. The Wollaston prism's splitting of ordinary and extraordinary rays introduces axial coma in the y-direction, which increases with the FOV. So the different routes for the trade study are proposed to be as follows:
\begin{itemize}
    \item Reducing and optimizing the FOV for the polarimetric mode of ComCam after adding the Wollaston prism
    \item Analyzing the split angle and imaging performance with a Wollaston prism made of different birefringent materials such as Quartz and MgF2.
    \item Designing a corrector lens as a part of the camera optics to correct for the axial coma
    \item Designing and optimizing for new camera optics that can be swapped in with the imaging optics for the polarimetric mode.
\end{itemize}
\section{Summary and Future work}
\label{sec5}
With their large collecting area, the next-generation Giant Segmented Mirror Telescopes (GSMTs) will enable the high-precision polarization measurements of fainter targets. This paper presents a brief list of polarimetric science cases and technical requirements for one of the GSMTs, the Giant Magellan Telescope. As polarization measurements are limited by IP and crosstalk, we have modeled the IP and crosstalk for GMT using the polarization ray tracing algorithm. We estimate the IP and crosstalk for GMT to be $<$0.05\% in the FOV of 4' wavelength range of 0.35-10 $\mu$m. The estimated IP and crosstalk are found to be within the requirements put forth by the science cases.
Further, we compare the IP and crosstalk at the Nasmyth focus for different GSMTs and find that GMT shows similar polarization effects as TMT and ELT at the Nasmyth focus. However, among the three GSMTs, GMT has an advantage of Gregorian focus where the instrumental polarization and crosstalk are $<$0.05\%, which could be a suitable location for the polarimeter. We investigate the possibility of incorporating polarimetry with the existing instruments for GMT and propose to perform a trade-off study with the Commissioning Camera (ComCam)
\appendix    
\acknowledgments 
The authors would like to thank Dr. Rebecca Bernstein (Giant Magellan Telescope project), Dr. Jeffrey Crane (PI: ComCam), and the entire team of ComCam for providing us with the optical design of ComCam.  The authors want to acknowledge the GMT-Pol team members for their contributions to the science cases and technical requirement document. \\ \\
\textbf{Members of GMT-Pol team}\\ 
\textit{Transients:} Grant Williams (University of Arizona), J. Craig Wheeler (University of Texas), Manisha Shreshta (University of Arizona),
Doug Leonard (San Diego State University), Jennifer Hoffman (University of Denver), Lifan Wang (Texas A\&M University), G C Anupama (Indian Institute of Astrophysics, India), Klaas Wiersema (University of Warwick) \\ \\
\textit{Stars and Their Environments:} B-G Andersson (SOFIA, NASA Ames), Hee-Won Lee (Sejong University, Korea), Seok Jun Chang (MPIA, Germany), Kostas Tassis (University of Crete, Greece), Nikos Mandarakas (University of Crete, Greece) \\ \\
\textit{Galaxies:} Ann Zabludoff (University of Arizona), Yujin Yang (KASI), Enrique Lopez Rodriguez (KIPAC, Stanford), Paul Smith (University of Arizona), Helen Jermak (Liverpool John Moores University) \\ \\
\textit{Solar System Objects:}Ludmilla Kolokolova (University of Maryland), and Padma Yanamandra-Fisher (Space Science Institute), Stefano Bagnulo (Armagh Observatory and Planetarium) \\ \\
\textit{Exoplanets and Circumstellar disks:} Christoph Keller (Lowell Observatory), Chris Packham (University of Texas), Ewan Douglas (University of Arizona), Maxwell Millar-Blanchaer (University of California, Santa Barbara), Ramya M. Anche (University of Arizona), and Daniel Cotton (Monterey Institute)

\bibliography{report} 
\bibliographystyle{spiebib} 

\end{document}